\documentclass[sigconf]{acmart}

\usepackage{soul}
\usepackage{url}
\usepackage{inputenc}
\usepackage{caption}
\usepackage{graphicx}
\usepackage{amsmath}
\usepackage{booktabs}
\usepackage{algorithm}
\usepackage{algorithmic}
\usepackage{stmaryrd}
\usepackage{amsfonts}
\usepackage{multirow}
\usepackage{multicol}
\usepackage{subfigure}
\usepackage{bbding}
\usepackage{pifont}
\usepackage{wasysym}
\usepackage{amssymb}
\usepackage{latexsym}
\usepackage{mathrsfs}
\usepackage{bm}
\usepackage{CJK}
\usepackage{geometry}
\usepackage{blindtext}
\usepackage{amsthm}
\usepackage{flushend}
\theoremstyle{definition}
\newtheorem{myDef}{Definition}

\newcommand{\our}{\textsc{HGANE}}
\newcommand{{\var}}{\textsc{ANE}}

\def\BibTeX{{\rm B\kern-.05em{\sc i\kern-.025em b}\kern-.08emT\kern-.1667em\lower.7ex\hbox{E}\kern-.125emX}}

\setcopyright{acmcopyright}
\begin{document}

\fancyhead{}
  
%


\begin{sloppypar}


\title{Collective Link Prediction Oriented Network Embedding \\ with Hierarchical Graph Attention}

\author{Yizhu Jiao}
\affiliation{\institution{Shanghai Key Laboratory of Data Science, Shanghai Institute for Advanced Communication and Data Science, \\ School of Computer Science, Fudan University, \\ Shanghai, China, yzjiao18@fudan.edu.cn}}
\email{}

\author{Yun Xiong}
\affiliation{\institution{Shanghai Key Laboratory of Data Science, Shanghai Institute for Advanced Communication and Data Science, \\  School of Computer Science, Fudan University, \\ Shanghai, China, yunx@fudan.edu.cn}}
\email{}

\author{Jiawei Zhang}
\affiliation{\institution{IFM Lab, Department of Computer Science \\ Florida State University, Tallahassee, FL, USA}}
\email{jiawei@ifmlab.org}

\author{Yangyong Zhu}
\affiliation{\institution{Shanghai Key Laboratory of Data Science, Shanghai Institute for Advanced Communication and Data Science, \\ School of Computer Science,  Fudan University, \\ Shanghai, China, yyzhu@fudan.edu.cn}}
\email{}

%
\renewcommand{\shortauthors}{Jiao and Xiong, et al.}

%
\begin{abstract}
To enjoy more social network services, users nowadays are usually involved in multiple online sites at the same time. Aligned social networks provide more information to alleviate the problem of data insufficiency. In this paper, we target on the collective link prediction problem and aim to predict both the intra-network social links as well as the inter-network anchor links across multiple aligned social networks. It is not an easy task, and the major challenges involve the network characteristic difference problem and different directivity properties of the social and anchor links to be predicted. To address the problem, we propose an application oriented network embedding framework, Hierarchical Graph Attention based Network Embedding ({\our}), for collective link prediction over directed aligned networks. Very different from the conventional general network embedding models, {\our} effectively incorporates the collective link prediction task objectives into consideration. It learns the representations of nodes by aggregating information from both the intra-network neighbors (connected by social links) and inter-network partners (connected by anchor links). What's more, we introduce a hierarchical graph attention mechanism for the intra-network neighbors and inter-network partners respectively, which resolves the network characteristic differences and the link directivity challenges effectively. Extensive experiments have been conducted on real-world aligned networks datasets to demonstrate that our model outperformed the state-of-the-art baseline methods in addressing the collective link prediction problem by a large margin. 
\end{abstract}

%
%


\copyrightyear{2019}
\acmYear{2019}
\setcopyright{acmcopyright}
\acmConference[CIKM '19] {The 28th ACM International Conference on Information and Knowledge Management}{November 3--7, 2019}{Beijing, China}
\acmPrice{15.00}
\acmDOI{10.1145/3357384.3357990}
\acmISBN{978-1-4503-6976-3/19/11} 


%

%
\maketitle

\section{Introduction}

Nowadays, online social networks have become very popular and extensively used in our lives. To enjoy more services, it is ubiquitous for users to participate in multiple online social platforms concurrently. For example, users may share photos with Instagram and check the latest news information via Twitter. To simplify the sign up/in process, most social platforms usually allow users to use their existing Twitter/Facebook/Google IDs to create their accounts at these new social sites, which will align different online networks together naturally. Each of these platforms can be represented as a massive network where nodes represent user accounts and intra-network links represent the social relationships among users. Especially, accounts owned by the same user in different networks are defined as anchor nodes \cite{kong2013inferring} and inter-network corresponding relationships between the anchor users are defined as anchor links \cite{kong2013inferring}. Different online networks connected by anchor links are defined as multiple aligned social networks \cite{zhang2014meta}.

In recent years, there has been a surge of interest in multi-network analysis. Traditional methods that target on one single network require sufficient information to build effective models. However, as proposed in  \cite{zhang2013predicting}, this assumption can be violated seriously when dealing with the cold start \cite{leroy2010cold} and data sparsity problems. The study of multiple aligned networks provides a direction to alleviate the data insufficiency problem. Some research works propose to transfer information across networks by anchor links to enhance the link prediction results within multiple networks mutually \cite{cao2018neural, zhang2017bl, zhang2014meta}. Besides, many existing works aim at anchor link formation prediction \cite{wang2018user, li2018matching, liu2016aligning}. However, most of these works study either intra-network or inter-network link prediction tasks separately. As discovered in \cite{zhang2014transferring}, multiple link prediction tasks in the same networks can actually be done simultaneously and enhanced mutually due to their strong correlations. 

\begin{figure}[!tbp]
     \centering
     \includegraphics[width=0.7\linewidth]{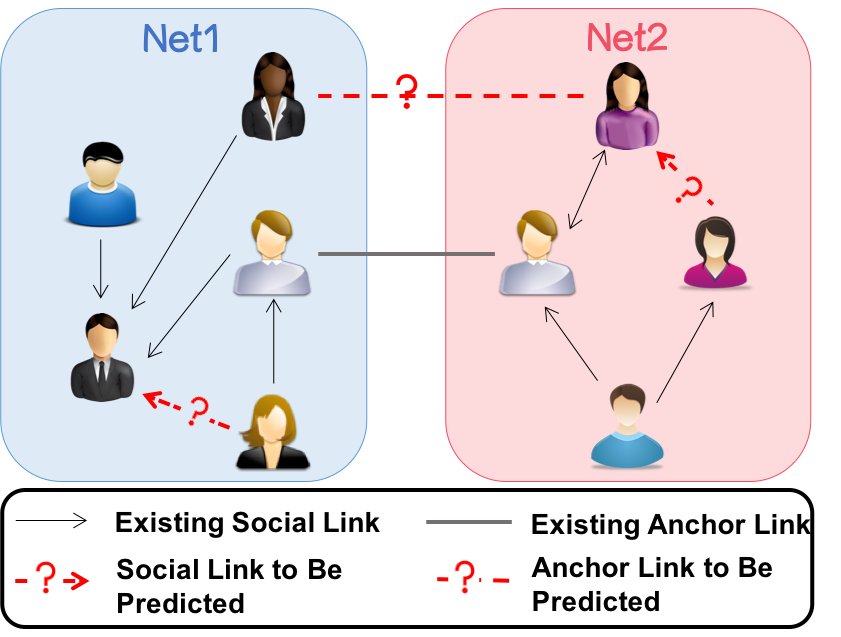}
     \caption{An example of collective link prediction over multiple aligned networks}
     \label{fig:task}
     \vspace{-10pt}
 \end{figure}

Predicting multiple kinds of links among users across multiple aligned networks is defined as the collective link prediction problem in  \cite{zhan2018integrated}. The collective link prediction problem covers several different link formation prediction tasks simultaneously including both the intra-network social link prediction and the inter-network anchor link prediction. It can take advantage of the strong correlations between these different tasks to enhance the prediction performance across these aligned networks synergistically. Figure \ref{fig:task} shows an example of the collective link prediction tasks of two social networks. In the figure, black lines with the arrow indicate existing directed intra-network social links and the gray lines indicate the existing inter-network anchor links instead. These directed/undirected red dashed lines with question marks signify the potential intra-network and inter-network links to be predicted, respectively. 

The problem of collective link prediction is worth exploring due to both its importance and novelty. Some existing methods have been introduced to tentatively address the problem \cite{zhan2018integrated}. However, these existing methods mostly ignore the contradiction of different characteristics of aligned networks or adopt fixed parameters to control the proportion of information diffused across networks, which usually need to be fine-tuned manually. Besides, these works also fail to consider the connectivity of the links within and across networks. 

The collective link prediction problem studied in this paper is also very challenging to solve due to the following reasons:
\begin{itemize}
\item \textbf{Network characteristic differences}: Since users normally join in different networks for different purposes, each network usually has different characteristics and reveals different aspects of the users. For example, professional relations are established on LinkedIn while personal social-tiers are built in Twitter. Thus, information transferred from other networks may be different from the target network that we want to study. Previous work found information transfer could also deteriorate the performance of intra-network link prediction \cite{zhang2014meta}. Correspondingly, anchor link prediction can be more susceptible as anchor links are directly related to information transferred across networks. Therefore, it is more crucial but challenging to overcome network characteristic difference problem for collective link prediction. 
\item \textbf{Link directivity differences}: The intra-network social links are usually uni-directed from the initiator pointing to the recipient instead. For the users involved in the social network, the social links pointing to them reflect the objective recognition from the community, whereas that from them to others reflect their personal social interest. Thus, these social relations collaboratively define a unique character in social networks. However, the inter-network anchor links are bi-directed according to the definition. Such different directivity properties on social links and anchor links should be carefully considered in the prediction model.
\end{itemize}

In this paper, we propose a novel application oriented network embedding framework, namely Hierarchical Graph Attention based Network Embedding ({\our}), to solve the collective link prediction problem over aligned networks. Very different from the conventional general network embedding models, {\our} effectively incorporates the collective link prediction task objectives into consideration. It learns node embeddings in multiple aligned networks by aggregating information from the related nodes, including both the intra-network social neighbors and inter-network anchor partners. What's more, we introduce a hierarchical graph attention mechanism for the intra-network neighbors and inter-network partners respectively, which handles the network characteristic differences and link directivity differences. {\our} balances between the prediction tasks of both the intra-network social link and inter-network anchor link and the learned embedding results can resolve the collective link prediction problem effectively. We conduct detailed empirical evaluations using several real-world datasets and show that our model outperforms other competitive approaches. 

We summarize the main contributions of this paper as follows:
\begin{itemize}
  \item We propose a novel embedding framework to learn the representations of nodes by aggregating information from both the intra-network neighbors (connected by social links) and inter-network partners (connected by anchor links).
  \item We introduce a hierarchical graph attention mechanism. It includes two levels of attention mechanisms — one at the node level and one at the network level —  to resolve the network characteristic differences and link directivity challenges effectively. 
  \item {\our} incorporates the collective link prediction task objectives into consideration and balances between the prediction tasks of the intra-network social link and inter-network anchor link, respectively.
  \item Extensive experiments are conducted on two real-world aligned social network datasets. The results demonstrate that the proposed model outperforms existing state-of-the-art approaches by a large margin.
\end{itemize}

\section{Preliminary}

\begin{myDef} (Multiple Aligned Social Networks) : In this paper, we follow the definitions introduced in \cite{kong2013inferring}. Given $n$ networks $\{G^{(1)}, \ldots, G^{(n)}\}$ with shared users, they can be defined as multiple aligned networks $\mathcal{G}=((G^{(1)}, G^{(2)}, \cdots, G^{(n)}) $,  $(A^{(1, 2)}, A^{(1, 3)}, \cdots, A^{(1, n)}, A^{(2, 3)}, \cdots, A^{(n - 1, n)}))$, where $ G^{(i)}= (V^{(i)}, E^{(i)}), i \in \{1, 2, \cdots, n\}$ is a network consisting of nodes and links, and $A^{(i, j)}$ represents the anchor links between $G^{(i)}$ and $G^{(j)}$. 
\end{myDef}

For two nodes $v^{(i)}\in V^{(i)}$ and $v^{(j)}\in V^{(j)}$,  node pair $(v^{(i)}, v^{(j)}) \in A^{(i, j)}$ iff $v^{(i)}$ and $v^{(j)}$ are the accounts of the same user in networks $G^{(i)}$ and $G^{(j)}$ respectively.

For two online networks, such as Foursquare and Twitter used in this paper, we can represent them as two aligned social networks $\mathcal {G}= ((G^{(1)}, G^{(2)}),(A^{(1, 2)}))$, which will be used as an example to illustrate the model. A simple extension of the proposed framework can be applied to multiple aligned networks conveniently.\\

{\bf Problem Definition} : The collective link prediction problem studied in this paper includes simultaneous prediction of both intra-network social links and inter-network anchor links. Formally, given two aligned networks $\mathcal {G}= ((G^{(1)}, G^{(2)}), (A^{(1, 2)}))$ where both of $G^{(1)}$ and $G^{(2)}$ are directed social networks. We can represent all the unknown social links among the nodes in $G^{(k)}$ as  $ U^{(k)} = V^{(k)} \times V^{(k)} \backslash (E^{(k)} \cup \{(u, u)\}_{u \in V^{(k)}}) $ where $k \in \{1,2\}$. And the unknown anchor links across $G^{(1)}$ and $G^{(2)}$ can be denoted as $ U^{(1,2)} = V^{(1)} \times V^{(2)} \backslash A^{(1,2)} $. We aim at building a mapping $f : U^{(1)} \cup U^{(2)} \cup U^{(1,2)} \rightarrow [0, 1]$ to project these intra-network social links and inter-network anchor links to their formation probabilities.

\section{Proposed method}

\begin{table}[!htbp]
\center
\caption{Descriptions of notations in our framework.}
\label{tab:notation}
\begin{tabular}{c|c}
\toprule
  Notation & Description \\
\midrule
  $u_i$ & Node i in $G^{(1)}$ \\
  $v_j$ & Node j in $G^{(2)}$ \\
  $\mathbf{u}_i^{in}$ & Initiator feature of $u_i$ in $G^{(1)}$ \\
  $\mathbf{v}_j^{in}$ & Initiator feature of $v_i$ in $G^{(2)}$ \\
  $\mathbf{u}_i^{re}$ & Recipient feature of $u_i$ in $G^{(1)}$ \\
  $\mathbf{v}_j^{re}$ & Recipient feature of $v_i$ in $G^{(2)}$ \\
  $\mathcal{N}^{i}(u_i) $ & Intra-network neighbors of $u_i$ as the initiator\\
  $\mathcal{N}^{r}(u_i) $ & Intra-network neighbors of $u_i$ as the recipient\\
  $\mathcal{N}^{a}(u_i)$ & Inter-network anchor partners of $u_i$\\
  $\mathbf{e}^{in}(u_i, u_j)$ & Intra-network initiator attention of $u_i$ to $u_j$ \\
  $\mathbf{e}^{in}(u_i, v_j)$ & Inter-network initiator attention of $u_i$ to $v_j$ \\
  $\mathbf{e}^{re}(u_i, u_j)$ & Intra-network recipient attention of $u_i$ to $u_j$ \\
  $\mathbf{e}^{re}(u_i, v_j)$ & Inter-network recipient attention of $u_i$ to $v_j$ \\
  \bottomrule
\end{tabular}
\end{table}

In this section, we will introduce the framework {\our} in detail. For the convenience of elaboration, we provide the main notations used through this paper in Table \ref{tab:notation}. At the beginning, the hierarchical graph attention mechanism will be introduced to handle the problems of network characteristic differences and link directivity challenges. After that, we will propose a novel node embedding method in multiple aligned networks. Finally, we will introduce the application oriented network embedding framework which can resolve the collective link prediction problem effectively.

\subsection{Hierarchical Graph Attention Mechanism} 
 \begin{figure}[!htbp]
     \centering
     \includegraphics[width=0.8\linewidth]{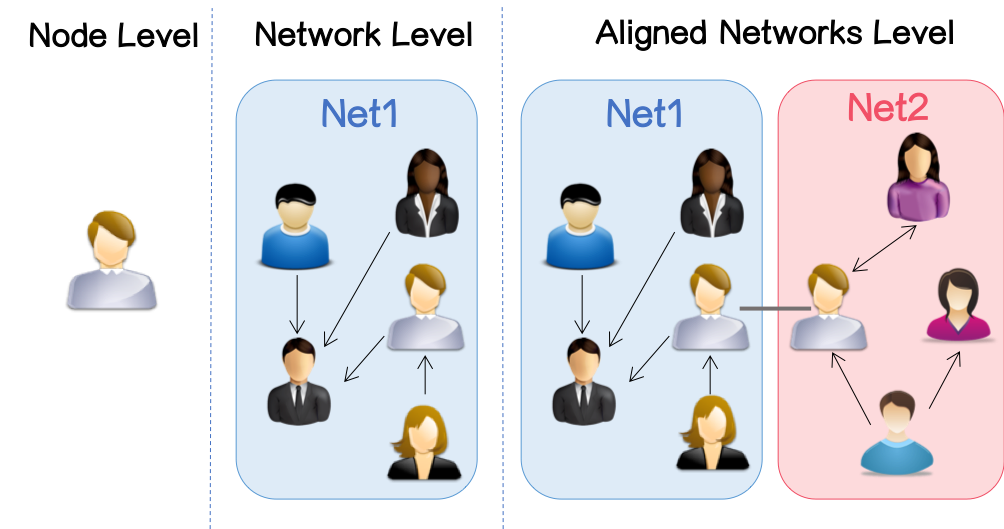}
     \caption{Hierarchical structure of multiple aligned networks}
     \label{fig:Hierarchical}
 \end{figure}
 
Social networks consist of nodes and social links, while multiple aligned social networks consist of many social networks and anchor links across them. Therefore, multiple aligned social networks have a hierarchical structure, which is illustrated in Figure \ref{fig:Hierarchical}. Besides, for the target node, it is observed that the relevance of different neighbors is different. For the target network, other networks are differentially informative since they have different characteristics. Furthermore, as each node is cooperatively characterized by its neighbor nodes and anchor partners in other networks, their importance is highly dependent on node embeddings.

Therefore, we propose the hierarchical graph attention mechanism in this section. It includes two levels of attention mechanisms  — one at the node level and the other at the network level  — to make our model pay more or less attention to different neighbor nodes and networks when constructing the node representations. These two levels of attention mechanisms are formally called the \emph{intra-network social attention} and the \emph{inter-network anchor attention}. They are essential to resolve the problems of network characteristic differences and link directivity challenges in the multiple aligned networks. In the following subsections, we will introduce the definitions and motivations of these two attention mechanisms.

\subsubsection{Intra-Network Social Attention} 

\ 

Given two aligned networks $\mathcal {G}= ((G^{(1)}, G^{(2)}),(A^{(1, 2)}))$, the anchor links are defined to be bi-directed; while the intra-network social links are usually uni-directed from the initiator pointing to the recipient instead. Thus, every node plays these two roles within the social network. Normally, we represent each node $u_i$ with two vector representations, the initiator feature $\mathbf{u}_i^{in} \in \mathbb R^d$ and the recipient feature $\mathbf{u}_i^{re} \in \mathbb R^d$, where $d$ is the feature dimension. The initiator feature $\mathbf{u}_i^{in}$ represents the characteristic of the node as the initiator following others while the recipient feature $\mathbf{u}_i^{re}$ represents the characteristic of the node as the recipient followed by others. By distinguishing the initiator and recipient features for each node, we can lay the foundation for resolving the problem of link directivity challenges effectively. 

For the target node $u_i$, we define the nodes followed by it as its intra-network recipient neighbors. The set of intra-network recipient neighbors of the initiator $u_i$ is denoted as $\mathcal{N}^{i}(u_i)$. The node $u_j \in \mathcal{N}^{i}(u_i)$ iff $(u_i, u_j) \in E^{(1)}$. Similarly, the nodes following $u_i$ are defined as its intra-network initiator neighbors and the set of these neighbors of the recipient $u_i$ is denoted as $\mathcal{N}^{r}(u_i)$. The node $u_j \in \mathcal{N}^{r}(u_i)$ iff $(u_j, u_i) \in E^{(1)}$. Here, based on the two node roles, we introduce \emph{intra-network initiator attention} and \emph{intra-network recipient attention}, to leverage the structural information within the social network.

For intra-network social neighbors, the characteristic of the initiator is relevant to the recipient. The intra-network initiator attention mechanism computes the coefficients to judge the importance of the intra-network recipient neighbor to the target initiator. Here, the concept of intra-network initiator attention mechanism can be represented formally. 

\begin{myDef}
  (Intra-Network Initiator Attention) :  For the target node $u_i$ and its intra-network recipient neighbor $u_j \in \mathcal{N}^{i}(u_i)$, the intra-network initiator attention coefficient of $u_i$ to $u_j$ can be given as
	\begin{displaymath}
    		 e^{in}(u_i, u_j) = \sigma \Big({\mathbf a_{in}^{(1)^T} } \Big[ \mathbf W_{in}^{(1)} {\mathbf u_i^{in}}\Big\| \mathbf W_{re}^{(1)} \mathbf u_j^{re} \Big] \Big)   , \nonumber 
	\end{displaymath}
where $\cdot^T$ represents transposition and $\|$ is the concatenation operation. ${\mathbf W_{in}^{(1)}}\in \mathbb R^{d^{\prime}\times d}$ and ${\mathbf W_{re}^{(1)}}\in \mathbb R^{d^{\prime}\times d}$ are the weight matrixes applied to every node as the initiator and the recipient for shared linear transformation in $G^{(1)}$. ${\mathbf a_{in}^{(1)^T} } \in \mathbb{R}^{2 d^{\prime}}$ is a weight vector and $\sigma$ denotes the activation function. ${\mathbf a_{re}^{(1)^T} }$,  ${\mathbf a_{in}^{(1, 2)^T}}$ and ${\mathbf a_{re}^{(1, 2)^T}}$ will be used with similar meanings. 
\end{myDef}

Similarly, since the characteristics of the initiator and the recipient are correlative, we can introduce the definition of intra-network initiator attention mechanism formally to obtain the importance of the intra-network initiator neighbor to the target recipient. 
\begin{myDef}
  (Intra-Network Recipient Attention) :  For the target node $u_i$ and its intra-network initiator neighbor $u_j \in \mathcal{N}^{r}(u_i)$, the intra-network recipient attention coefficient of $u_i$ to $u_j$ can be given as
	\begin{displaymath}
   		e^{re}(u_i, u_j) = \sigma \Big({\mathbf a_{re}^{(1)^T} } \Big[ \mathbf W_{re}^{(1)} {\mathbf{u}_i^{re}}\Big\| \mathbf W_{in}^{(1)} \mathbf{u}_j^{in}\Big] \Big) \nonumber, 
	\end{displaymath}
where ${\mathbf a_{re}^{(1)^T} }$ is also a weight vector. 
\end{myDef}
With these two kinds of intra-network attention mechanism, our model can pay more attention on useful information and neglect harmful information within the social networks. It is significant to resolve the problem of directivity challenges effectively and leverage structural information within social networks. 

\subsubsection{Inter-Network Anchor Attention}

\ 

Different from the intra-network social attention mechanism which targets at the node level, the inter-network anchor attention is for the network level. The anchor links connecting multiple networks play a crucial role in cross-network information transfer. However, due to the problem of network characteristic differences, information transferred from other networks could also undermine the learned embeddings of the target network. 

To handle this problem, we introduce the inter-network anchor attention mechanism. For the target node, the inter-network anchor attention coefficient to its anchor partners in the other network indicates the importance of information transferred from that network. As mentioned in the last section, each node is represented by the initiator and recipient embeddings. To transfer the directed structural information within networks, two kinds of inter-network anchor attention will be introduced according to these two roles of the nodes. 

Different from uni-directed social links, since the anchor nodes reveal the information of the same user from different aspects, it is intuitive that their initiator and recipient features in different networks can be related correspondingly. For the target node and its anchor partner in some network, the inter-network initiator attention coefficient indicates the importance of information from that network to the target node as the initiator. Firstly, we represent the set of the inter-network anchor partners in the other network for the target node $u_i$ as $\mathcal{N}^{a}(u_i)$. If the node pair $(u_i, v_j) \in A^{(1, 2)}$, $v_j \in \mathcal{N}^{a}(u_i)$. And the definition of inter-network initiator attention is introduced as follows.

\begin{myDef}
  (Inter-Network Initiator Attention): For the target node $u_i$ and its inter-network anchor partner $v_j \in \mathcal{N}^{a}(u_i)$, the intra-network initiator attention coefficient of $u_i$ to $v_j$ as the initiator can be given as
\begin{displaymath}
    e^{in}(u_i, v_j) = \sigma \Big({\mathbf a_{in}^{(1, 2)^T}} \Big[ \mathbf W_{in}^{(1)} {\mathbf{u}_i^{in}}\Big\| \mathbf W_{in}^{(1, 2)} \mathbf{v}_j^{in}\Big] \Big) \nonumber, 
\end{displaymath}
where $\mathbf W_{in}^{(1, 2)}$ is the weight matrix applied to every anchor node in $G^{(2)}$ as the initiator while transferring information to $G^{(1)}$ and ${\mathbf a_{in}^{(1, 2)^T}}$ is a weight vector.
\end{myDef}

Similarly, considering the recipient role of nodes, it is intuitive that the recipient features of anchor partners are related. Based on this, we give the concept of inter-network recipient attention, which denotes the importance of information from that network to the target node as the recipient.

\begin{myDef}
  (Inter-Network Recipient Attention) :  For the target node $u_i$ and its inter-network anchor partner $v_j \in \mathcal{N}^{a}(u_i)$, the intra-network recipient attention coefficient of $u_i$ to $v_j$ as the recipient can be given as
\begin{displaymath}
    e^{re}(u_i, v_j) =  \sigma \Big({\mathbf a_{re}^{(1, 2)^T}} \Big[ \mathbf W_{re}^{(1)} {\mathbf{u}_i^{re}}\Big\| \mathbf W_{re}^{(1, 2)} \mathbf{v}_j^{re}\Big] \Big) \nonumber, 
\end{displaymath}
where $\mathbf W_{re}^{(1, 2)}$ is the weight matrix for every anchor node in $G^{(2)}$ as the recipient and ${\mathbf a_{re}^{(1, 2)^T}}$ is also a weight vector.
\end{myDef}

Inter-network anchor attention mechanism can make a great contribution to effective cross-network information transfer. It handles the problem of network characteristic differences by making our model focus on more important networks with useful information. 


Besides, to make multiple coefficients easily comparable across different nodes, we normalize all the coefficients mentioned above across all choices of intra-network neighbors and inter-network partners using the softmax function. Thus, the four kinds of attention coefficients of the concerned nodes within networks and across networks can be rewritten into a unified formula: 
\begin{align}
    \alpha^{op}(u_i, u_j/v_j) &= \emph{softmax}_{u_j/v_j}(e^{op}(u_i, u_j/v_j)) \nonumber \\
    &= \exp({e^{op}(u_i, u_j/v_j) )} \Big/ SUM_{u_i}^{op} \nonumber
\end{align}
, where $op \in \{in, re\}$ indicates the role of the target node. And the denominator is set as
\begin{align}
    SUM_{u_i}^{op} = {\sum_{u_k \in \mathcal{N}^{op}(u_i)} \exp (\mathbf{e}^{op}(u_i, u_k))} + {\sum_{v_k \in {\mathcal{N}^{a}(u_i)}}\exp (\mathbf{e}^{op}(u_i, v_k))}\nonumber 
\end{align}

\subsection{Multiple Aligned Network Embedding} 
 \begin{figure*}[!htbp]
     \centering
     \includegraphics[width=0.9\textwidth]{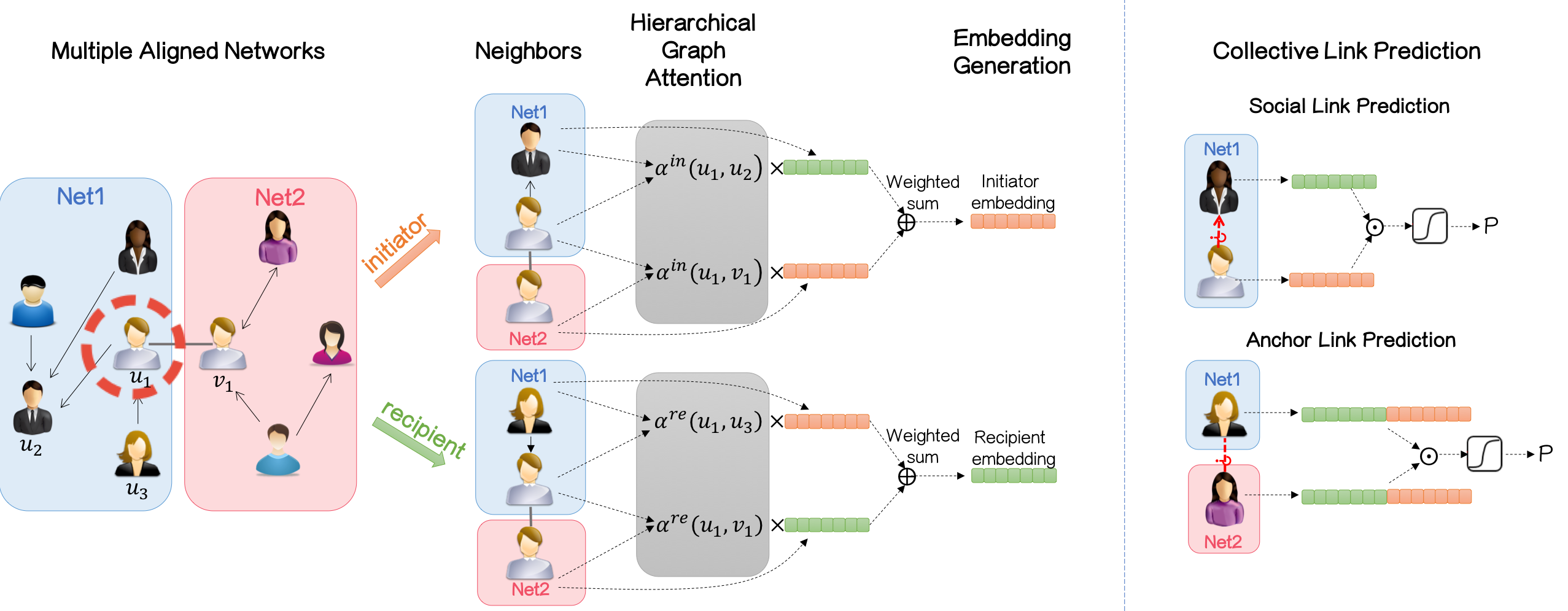}
     \caption{{\our} model architecture. Each node is represented by the initiator and recipient embeddings (the orange and green little squares) in the aligned networks. The left plot provides the example of learning node embeddings ($u_1$ in Net1) by aggregating information from both the intra-network neighbors ($u_2$, $u_3$ in Net1) and inter-network partners ($v_1$ in Net2), which is weighted by hierarchical graph attention. The right plot illustrates how {\our} resolves the collective link prediction with the learned embeddings.}
     \label{fig:model}
 \end{figure*}
With the hierarchical graph attention introduced in the previous section, we will introduce the cross-network embedding framework {\our} in this part. {\our} is based on the cross-network graph neural network model, which extends the traditional graph neural network (GNN) model \cite{scarselli2009graph} to the multiple aligned social networks scenario. According to the principle of GNN, embeddings can capture the localized structural features by utilizing information propagated from the intra-network social neighbors. What's more, it can preserve more comprehensive features by leveraging cross-network information transferred by anchor links. Therefore, {\our} learns the node representations by aggregating information from both the intra-network neighbors and inter-network partners. At the same time, {\our} takes advantage of the hierarchical graph attention mechanism to focus on more important information to handle the problems of network characteristic differences and link directivity challenges. The architecture of {\our} is illustrated in Figure \ref{fig:model}.

Each node is represented by two embeddings, $\mathbf{u}_i^{in}$ and $\mathbf{u}_i^{re}$, according to its two role in social networks. The implicit initiator and recipient representations of the node are represented as ${\mathbf{u}_i^{in}}^{\prime} \in \mathbb{R}^{d^{\prime}}$ and ${\mathbf{u}_i^{re}}^{\prime}\in \mathbb{R}^{d^{\prime}}$ of the potentially different dimension $d^\prime$.

The initiator embedding of $u_i$, $\mathbf{u}_i^{in}$, which indicates its features as the initiator in the social network, depends on its intra-network recipient neighbors. Therefore, for the node $u_j \in \mathcal{N}^{i}(u_i)$, $\mathbf{u}_j^{re}$ can contribute to $\mathbf{u}_i^{in}$ with the coefficient, $\alpha^{in}(u_i, u_j)$, determined by intra-network initiator attention. We define the \emph{intra-network neighbor recipient contribution} (NRC) from $u_j$ to $u_i$  as 
\begin{align}
NRC(u_i, u_j) = \mathbf{\alpha}^{in}(u_i, u_j) \mathbf W_{re}^{(1)} \mathbf{u}_j^{re} \nonumber
\end{align}

As to inter-network anchor partners, node embeddings can preserve more comprehensive information by taking their features of the same role in other networks into consideration. For the anchor node $ v_j \in \mathcal{N}^{a}(u_i)$, its initiator embedding $\mathbf{v}_j^{in}$ also contribute to the initiator embedding $\mathbf{u}_i^{in}$ of the target node. To overcome the problem of network characteristic differences, the inter-network initiator attention will compute the weights of information from different networks. The \emph{inter-network partner initiator contribution} (PIC) from $v_j$ to $u_i$  is introduced as 
\begin{align}
PIC(u_i, v_j) = \mathbf{\alpha}^{in}(u_i, v_j)\mathbf W_{in}^{(1, 2)}\mathbf{v}_j^{in} \nonumber
\end{align}

Formally, we can obtain the initiator embedding of $u_i$ by aggregating the intra-network neighbor recipient contribution and the inter-network partner initiator contribution as

\begin{align}
     {\mathbf{u}_i^{in}}^{\prime} = \sigma  \Big({\sum_{{u_j} \in \mathcal{N}^{i}(u_i)}} NRC(u_i, u_j) + \sum_{{v_j} \in \mathcal{N}^{a}(u_i)} PIC(u_i, v_j) \Big), \nonumber
\end{align}

where $\sigma$ denotes a nonlinearity. Besides, the recipient embedding of $u_i$, $\mathbf{u}_i^{re}$, can be generated in the similar way. Inter-network and inter-network recipient attention will determine the importance of different related nodes' contribution. As the recipient, $u_i$ is naturally characterized by its intra-network initiator neighbors who actively follow it in social networks. Such intuition leads to the contribution from every node, such as $u_j \in \mathcal{N}^{r}(u_i)$, to $u_i$. The \emph{intra-network neighbor initiator contribution} (NIC) from $u_j$ to $u_i$  can be defined as 
\begin{align}
NIC(u_i, u_j) = \mathbf{\alpha}^{re}(u_i, u_j) \mathbf W_{in}^{(1)} \mathbf{u}_j^{in} \nonumber
\end{align}

The inter-network anchor partners provide more information about the recipient role of the target node from different sources. Thus, the recipient embedding of $u_i$ will aggregate the information from other networks by the anchor nodes with different weights computed by the inter-network recipient attention in other networks. We can obtain the \emph{inter-network partner recipient contribution} (PRC)  from $u_j$ to $u_i$  to be 
\begin{align}
PRC(u_i, v_j) = \mathbf{\alpha}^{re}(u_i, v_j)\mathbf W_{re}^{(1, 2)}\mathbf{v}_j^{re} \nonumber
\end{align}

By combining these contributions from intra-network initiator neighbors and inter-network partners, the recipient embedding of $u_i$ can be represented formally as
\begin{align}
     {\mathbf{u}_i^{re}}^{\prime} = \sigma \Big({\sum_{{u_j} \in \mathcal{N}^{r}(u_i)}} NIC(u_i, u_j) + \sum_{{v_j} \in \mathcal{N}^{a}(u_i)}  PRC(u_i, v_j) \Big) \nonumber
\end{align}

To stabilize the learning process of node embeddings, we have found extending our mechanism to employ multi-head attention to be beneficial, inspired by Vaswani et al. 2017 \cite{DBLP:journals/corr/abs-1710-10903}. Specifically, $K$ independent attention mechanisms execute the above transformation, and then their embeddings are concatenated, resulting in the following initiator and recipient feature representations:
\begin{align}
      {\mathbf{u}_i^{in}}^{\prime} = \bigparallel_{k = 1}^{K}\sigma \Big({\sum_{{u_j} \in \mathcal{N}^{i}(u_i)}} NRC_k(u_i, u_j) + \sum_{{v_j} \in \mathcal{N}^{a}(u_i)} PIC_k(u_i, v_j) \Big) \nonumber 
\end{align}
\begin{align}
      {\mathbf{u}_i^{re}}^{\prime} = \bigparallel_{k = 1}^{K}\sigma \Big({\sum_{{u_k} \in \mathcal{N}^{r}(u_i)}} NIC_k(u_i, u_j) + \sum_{{v_j} \in \mathcal{N}^{a}(u_i)}  PRC_k(u_i, v_j) \Big)  \nonumber 
\end{align}
where NRC, PIC, NIC and PRC with the subscript $k$ in the formulas denote the contributions computed with the $k$-th hierarchical attention mechanism.

With reference to above equations, the formula derivation of computing the node embeddings for $G^{(2)}$ can be obtained in the similar way. They are not listed due to the page limit. 


\subsection{Collective Link Prediction Oriented Network Embedding Framework}
The embeddings of each node in multiple aligned networks can be generated by aggregating information from both the intra-network neighbors and inter-network partners as introduced in last section. In this part, we will introduce the network embedding optimization framework oriented to collective link prediction. The task includes the simultaneous prediction of the social links within each network and the anchor links between every two networks. 

For a node pair $(u_i, u_j)$ within the social network, we define \emph{the probability of the intra-network social link formation} from the initiator $u_i$ pointing to the recipient $u_j$ as 
\begin{align}
    p\big(u_i, u_j\big) = \sigma \big({{\mathbf{u}_i^{in}}^{T} \cdot \mathbf{u}_j^{re}} \big), \nonumber
\end{align}%
where $\sigma(x) = 1/(1 + \exp(-x))$ is the sigmoid function. And we adopt the approach of negative sampling \cite{DBLP:journals/corr/abs-1301-3781} to define the objective of intra-network social link formation from the initiator $u_j$ to the recipient $u_i$ as 
\begin{align}
    \mathcal{L}_{soc}\big(u_i, u_j\big)  = \log p\big(u_i, u_j\big) + \sum_{\{(u_m, u_n)\}}\log \big(1 - p\big(u_m, u_n\big) \big),  \nonumber 
\end{align}%
where $\{(u_m, u_n)\}$ denotes the set of the negative social links random sampled from the unknown links among nodes in $G^{(1)}$. In the objective, the first term models the existing social links while the second term models the negative links. By adding the objective of each intra-network social link, the final objective for $G^{(1)} $ can be formally represented as
\begin{align}
    \mathcal L^{(1)} &= \sum_{(u_i, u_j) \in E^{(1)}} \mathcal{L}_{soc}(u_i, u_j) \nonumber
\end{align}%

Similarly, we can define the objective for the embedding results for $ G^{(2)} $, which can be formally represented as $\mathcal{L}^{(2)}$. 

Besides, anchor nodes reflect information of same users. Therefore, their features tend to be in a close region in the embedding space whether as the initiator or the initiator. For the cross-network node pair $(u_i, v_j)$ where $u_i \in E^{(1)}$ and $v_i \in E^{(2)}$, we concatenate the initiator and recipient embeddings of each node to define the \emph{the probability of the inter-network anchor link formation} as 
\begin{align}
    p\big(u_i, v_j\big) = \sigma \big(\big({{\mathbf{u}_i^{in}} \|{\mathbf{u}_i^{re}}}\big)^{T}\cdot  \big({{\mathbf{v}_j^{in}}\|{\mathbf{v}_j^{re}}}\big) \big)  \nonumber
\end{align}%

Similarly with the objective of the intra-network social link formation, the objective of node alignment with negative sampling can be defined as 
\begin{align}
    \mathcal{L}_{ach}\big(u_i, v_j\big) = \log p\big(u_i, v_j\big) + \sum_{\{(u_m, v_n)\}} \log \big(1 - p\big(u_m, v_n\big) \big), \nonumber
\end{align}%
where $\{(u_m, v_n)\}$ denotes the set of the negative anchor links random sampled from the unknown anchor links across $G^{(1)}$ and $G^{(2)}$. By aligning anchor nodes, we can leverage information from multiple sources to learn the node embeddings comprehensively. Information transfer across networks is achieved based on every inter-network anchor link. Formally, the information transfer objective between $ G^{(1)}$ and $ G^{(2)}$ is represented by summing up the the objective of each anchor link as 
\begin{align}
    \mathcal{L}^{(1, 2)} = \sum_{({u}_{i},{v}_{j})\in A^{(1, 2)}}\big(\mathcal{L}_{ach}\big(u_i, v_j\big)\big)\nonumber
\end{align}
To incorporate the collective link prediction task into a unified framework, we learn the node representations with rich information by jointly training the objective function including the objective for networks $ G^{(1)}$, $ G^{(2)}$, and the objective of information transfer, which can be denoted as
\begin{align}
    \mathcal{L}\left(G^{(1)}, G^{(2)}\right) = \mathcal{L}^{(1 )}+ \mathcal{L}^{(2)}+ \alpha \cdot \mathcal{L}^{(1,2)}+ \beta \cdot \mathcal{L}_{r e g}\nonumber
\end{align}%
The parameter $\alpha$ denotes the weight of the information transfer objective to balance between the several prediction tasks of both the intra-network social link and the inter-network anchor link. In the objective function, the term $\mathcal{L}_{reg}$ is added to avoid overfitting and the parameter $\beta$ denotes the weight of it. By optimizing the above objective function, the node embeddings can be learned to resolve the collective link prediction problem effectively.

\section{Experiment}
\begin{table}[!htbp]
\center
\caption{Statistics of datasets}
\label{tab:Datasets}
\begin{tabular}{c|c|c|c}
\toprule
Dataset & \#Nodes & \#Social Links & \#Anchor Links \\
\hline
Twitter & 5,223 & 164,920 & \multirow{2}*{3,388}\\
Foursquare & 5,392 & 76,972 \\
\hline
Facebook & 4,137 & 57,528 & \multirow{2}*{4,137}\\
Twitter & 4,137 & 147,726 \\
\bottomrule
\end{tabular}
\vspace{-10pt}
\end{table}

\subsection{Datasets}

We conducted experiments using two real-world aligned social networks: Twitter-Foursquare and Facebook-Twitter(Statistical information in Table \ref{tab:Datasets}):
\begin{itemize} 
\item {\bf Twitter-Foursquare} \cite{kong2013inferring}: Twitter is the most popular worldwide microblog site and Foursquare is the famous location-based social network. There are 5,223 users and 164,920 follow links in Twitter and 5,392 users and 76,972 social links in Foursquare. Among these crawled Foursquare users, 3,388 of them are aligned by anchor links with Twitter. 
\item {\bf Facebook-Twitter} \cite{cao2016asnets}: Facebook is another worldwide online social media. The Facebook and Twitter accounts of 4,137 users were crawled. Every node has a counterpart in the other network. There are 57,528 social links in Facebook and 147,726 follow links in Twitter among these users. 
\end{itemize}
It is noted that these two datasets were crawled respectively and there is no overlap of these two Twitter subnetworks.

\textbf{Source Code}: The source code of {\our} is available in \url{http://github.com/yzjiao/HierarchicalGraphAttention}.

\subsection{Comparison Methods}

\begin{table}[!htbp]
\center
\caption{Comparision of different models}
\label{tab:comparision}
\begin{tabular}{cccccc}
\toprule  
{\shortstack{Method\\ \qquad}}		&{\shortstack{Multiple\\Networks}}	&{\shortstack{Links\\Directi.\\Differe.}}	&{\shortstack{Network\\Charact.\\Differe.}}&{\shortstack{Predict\\Social\\Link}}	&{\shortstack{Predict\\Anchor\\Link}}\\       
\midrule
DeepWalk		&				&				&								&\Checkmark						&				\\
\hline
Node2Vec 	&				&				&								&\Checkmark						&				\\
\hline
GAT	 		&				&				&								&\Checkmark						&				\\
\hline
IONE    		&\Checkmark		&\Checkmark		&								& 								& \Checkmark		\\
\hline
DIME       		&\Checkmark		&\Checkmark		&								& \Checkmark						&				\\
\hline
MNN       		&\Checkmark		&\Checkmark 		&								&\Checkmark						&				\\
\hline
CLF    		&\Checkmark		&		 		&								&\Checkmark						& \Checkmark		\\
\hline
{\our} 	       &\Checkmark		&\Checkmark	 	&\Checkmark						& \Checkmark						& \Checkmark		\\
\bottomrule
\end{tabular}
\end{table}

\renewcommand\arraystretch{1}
\begin{table*}[!htbp]
\center
\caption{Performance comparison with different methods. Soc1, Soc2 and Ach indicate social link prediction in the first and second network and anchor link prediction respectively.}
\label{tab:result}
\begin{tabular}{c|c|ccc|ccc|ccc|ccc}
\toprule
\multirow{4}*{Dataset}& \multirow{4}*{Method}
& \multicolumn{12}{c}{Training Ratio}\\ 
\cmidrule(lr){3-14}
 & & \multicolumn{3}{|c|}{0.2}& \multicolumn{3}{|c|}{0.4}& \multicolumn{3}{|c|}{0.6}& \multicolumn{3}{|c}{0.8}\\ 
\cmidrule(lr){3-5}\cmidrule(lr){6-8}\cmidrule(lr){9-11}\cmidrule(lr){12-14}
 & & Soc1 & Soc2 & Ach & Soc1 & Soc2 & Ach & Soc1 & Soc2 & Ach & Soc1 & Soc2 & Ach  \\
\midrule
 \multirow{9}*{\shortstack{Twitter\\ \& \\Foursquare}}
  & DeepWalk 		&75.8\% 	& 72.5\% &57.9\%	&80.3\%	&76.9\%	& 63.1\%	&82.2\% 	&79.7\% 	& 67.3\%	&85.7\%	&82.5\%	& 75.4\%	\\
  & Node2Vec 		&82.5\%	& 77.4\%	&64.3\%	&84.6\%	&80.9\%	& 66.1\%	&86.4\% 	&84.3\%	& 72.1\%	&89.3\%	&88.3\%	& 78.9\% \\
  & GAT	 		&85.5\%	&78.2\%	&65.5\%  &91.5\%	&86.9\%	& 68.9\%  &92.5\%	&90.3\%	& 75.8\%  &92.6\%	&92.3\%	& 80.9\%  \\ 
  & IONE    		&83.2\%	&75.7\%	&72.1\%	&86.2\%	&81.7\%	& 78.0\%	&88.2\%	&84.7\%	& 85.6\%	&88.7\%	&84.7\%	& 87.4\%	\\
  & DIME       		&85.1\%	&76.2\% 	&74.8\%	&88.4\%	&80.3\%	& 76.3\%	&89.8\%	&83.0\%	& 82.6\%	&92.0\%	&85.2\%	& 84.9\%	\\
  & MNN       		&89.2\%	&72.4\% 	&-     	&92.9\%	&81.1\%	&  -     	&94.8\%	&86.1\%	&  - 		&96.3\%	&87.6\%	&  - 		\\
  & CLF    		&84.5\%	&78.7\%	&70.9\%	&86.7\%	&80.5\%	& 75.2\% &90.9\%	&84.2\%	& 83.1\%	&92.3\%	&86.5\%	& 87.1\%	\\
  & {\our} 			&{\bf 94.4\%}	&{\bf 90.3\%}	&{\bf 76.7\%}	&{\bf 96.4\%}	&{\bf 95.1\%}	&{\bf 85.8\%}	&{\bf 97.1\%}	&{\bf 96.8\%}	&{\bf 90.0\%}	&{\bf 97.5\%}	&{\bf 97.5\%}	&{\bf 93.0\%}	\\
 \cline{1-14}

 \multirow{9}*{\shortstack{Facebook \\ \& \\Twitter}}
  & DeepWalk 		&76.3\% 	& 70.3\% & 55.8\%	&81.5\%	&75.2\%	& 70.9\%	&84.0\% 	&81.6\% 	& 77.7\% 	&90.9\%	&86.5\%	&78.9\%	\\
  & Node2Vec 		&83.0\%	& 81.5\%	& 58.6\%	&86.6\%	&85.7\%	& 76.2\%	&88.8\% 	&87.5\%	& 81.0\%	&91.3\%	&88.2\%	&83.2\%	\\
  & GAT	 		&87.3\%	&86.1\%	& 60.2\% &92.0\%	&90.0\%	& 78.5\% &94.7\%	&92.8\%	& 83.5\% &95.7\%	&93.4\%	&85.5\%  \\ 
  & IONE    		&82.8\%	& 79.1\%	& 77.9\%	&85.9\%	&82.6\%	& 85.4\%	&87.4\%	& 85.1\%	& 89.4\%	&90.9\%	&89.1\%	&92.1\%	\\
  & DIME       		&87.1\%	&86.2\% 	& 74.3\%	&88.4\%	&87.3\%	& 81.9\%	&89.8\%	&90.0\%	& 85.1\% &94.0\%	&92.2\%	&87.5\%  \\
  & MNN       		&88.6\%	&87.1\% 	&-     	&92.4\%	&91.3\%	&  -     	&94.4\%	&93.1\%	&  - 		&95.7\%	&94.8\%	&  - 		\\
  & CLF    		&84.9\%	& 81.1\%	& 80.5\%	&88.7\%	&85.9\%	& 84.2\% &91.4\%	&88.9\%	& 87.6\%	&93.1\%	&90.2\%	&90.4\%	\\
  & {\our} 			&{\bf 91.8\%}	&{\bf 90.9\%}	&{\bf 84.8\%}	&{\bf 95.2\%}	&{\bf 94.8\%}	&{\bf 93.4\%}	&{\bf 97.1\%}	&{\bf 96.9\%}	&{\bf 95.8\%}	&{\bf 98.1\%}	&{\bf 97.5\%}	&{\bf 97.1\%}	\\
\bottomrule
\end{tabular}
\end{table*}

\renewcommand\arraystretch{1}
\begin{table}[!htbp]
\center
\caption{Validation of the design of represent each node with two embeddings to resolve the link directivity differences problem. Our full model outperforms two variants with either the initiator or recipient features. }
\label{tab:inputoutput}
\begin{tabular}{c|ccc|ccc}
\toprule
  \multirow{2.5}*{Feature}& \multicolumn{3}{|c|}{Twitter\&Foursquare}& \multicolumn{3}{|c}{Facebook\&Twitter}\\
  \cmidrule(lr){2-7}
  & Soc1 & Soc2 & Ach & Soc1 & Soc2 & Ach \\ 
\midrule
  initiator 	 &93.4\%	&93.2\%	&85.2\%	&97.0\%	&94.9\%	&95.8\%	\\
  recipient	 &93.0\% 	&93.7\% 	&85.6\%	&97.1\% 	&95.1\% 	&96.2\%	\\
  both	 &\bf{97.2\%}	&\bf{96.8\%}	&\bf{93.0\%}	&\bf{98.1\%}	&\bf{97.5\%}	&\bf{97.1\%}	\\
  \bottomrule
\end{tabular}
\vspace{-5pt}
\end{table}

The network embedding methods used in the experiment are listed as follows (summarized in Table \ref{tab:comparision}):
\begin{itemize}
\item {\bf DeepWalk} \cite{perozzi2014deepwalk}: Skip-gram based vertex embedding method for a single network that extends the word2vec \cite{mikolov2013distributed} to the network scenario.
\item {\bf Node2Vec} \cite{grover2016node2vec}: Word-to-vector approach for a single network that modifies the random walk strategy in DeepWalk into a more sophisticated schemes.
\item {\bf GAT} \cite{DBLP:journals/corr/abs-1710-10903}: A neural network architecture for a single network to learn node representation by leveraging masked self-attention layers.
\item {\bf IONE} \cite{liu2016aligning}: A representation learning model for multiple aligned network by preserving the proximity of users with “similar” followers/followees in the embedded space for network alignment.
\item {\bf DIME} \cite{zhang2017bl}: An embedding framework for multiple heterogeneous aligned network with aligned autoencoders to transfer information and improve the link prediction in emerging networks.
\item {\bf MNN} \cite{cao2018neural}: A multi-neural-network framework for intra-network link prediction over aligned networks. It is not suitable for anchor link prediction as it assigns anchor users with the same feature vectors. 
\item {\bf CLF} \cite{zhan2018integrated}: A method aiming at collective link prediction by propagating the probabilities of predicted links across the partially aligned networks with collective random walk.
\end{itemize}

\subsection{Experiment Setting}

In our experiment, we will target on the collective link prediction task and concern the performance of social link prediction in $G^{(1)}$, $G^{(2)}$, and anchor link prediction across these two networks. These three subtasks will be denoted as Soc1, Soc2, Ach in the experiment results. As link prediction is regarded as a binary classification task, the performance will be evaluated with Area Under the Curve (AUC) metric. 

All the existing links in the two aligned networks are used as the positive link set, including social links within two networks and anchor links across these two networks. We sample a subset of unknown links among nodes in the same network randomly as the negative social link set, which is of the double size of the positive social link set. The negative anchor link set is generated by the random sample of unknown cross-network links. The size of the negative anchor link set is five times of that of the positive set. A proportion of the links in the positive and negative sets are sampled as the training set, the rest as the test set.

For our embedding framework {\our}, we initialize the initiator and recipient features of each node with the common initiator and recipient neighbor features within its networks. There are two attention-based layers involved for each network. The first layer consists of K = 8 attention heads computing 256 features each, followed by an exponential linear unit (ELU) \cite{DBLP:journals/corr/ClevertUH15} nonlinearity. The second layer is a single attention head to compute node embeddings, followed by a softmax activation. The dimension of the embeddings is 100. During training, we apply dropout \cite{srivastava2014dropout} to the normalized attention coefficients. And we train for 3000 epochs using the Adam optimizer \cite{kingma2014adam} with the learning rate of 0.005. The parameters $\alpha$ = 1.0 and $\beta$ = 0.0005 are used in the experiments. 

For the comparison methods that target at one single network, such as Node2vec, DeepWalk and GAT, we preprocess the datasets by merging two networks into one and regarding anchor links as social links within networks. We apply the linear SVM classifier for those embedding methods that can't directly predict the formation of links. Notably, we have chosen optimal hyper-parameters carefully for different baselines in this paper to ensure the fairness of comparison experiments.

\subsection{Experiment Result}

In the collective link prediction task, we compare the performance of eight different embedding methods under different training rate $\lambda \in \left\{20\%, 40\%, 60\%, 80\% \right\}$. Table \ref{tab:result} shows the performance of our model and other seven baseline methods evaluated by AUC with different training rate $\lambda$. The method we proposed in this paper, {\our}, performs much better than the other methods in the three subtasks simultaneously, which shows its effectiveness in the collective link prediction task. {\our} incorporates the task-oriented objectives into consideration and thus balance between the prediction tasks of both the intra-network social link and inter-network anchor link simultaneously.  

Considering the experiments with different training rate $\lambda$, as the ratio drops, the performance of all the methods deteriorates. However, the performance degradation of the proposed model is rather moderate compared to other methods since we leverage the information of the multiple aligned networks and handle the problem of network characteristic differences. Even when the training rate $\lambda$ is as low as 20\%, the baseline models will suffer from the information sparsity a lot, but our model can still obtain very good performance. 

To demonstrate the effectiveness of considering the directivity, we compare our full model to its two variants with either the initiator features or the recipient features. The experiment results show in Table \ref{tab:inputoutput} with the training ratio as 0.8. We found that the two variants can give better results than other baselines but their performance is much inferior to that of the full model. According to the experimental statistics of two datasets, the performance of social link prediction within the more dense network can be improved more by distinguishing nodes' initiator and recipient roles.

\subsection{Hypothesis Verification}

\begin{figure}
     \centering
     \includegraphics[width=0.8\linewidth]{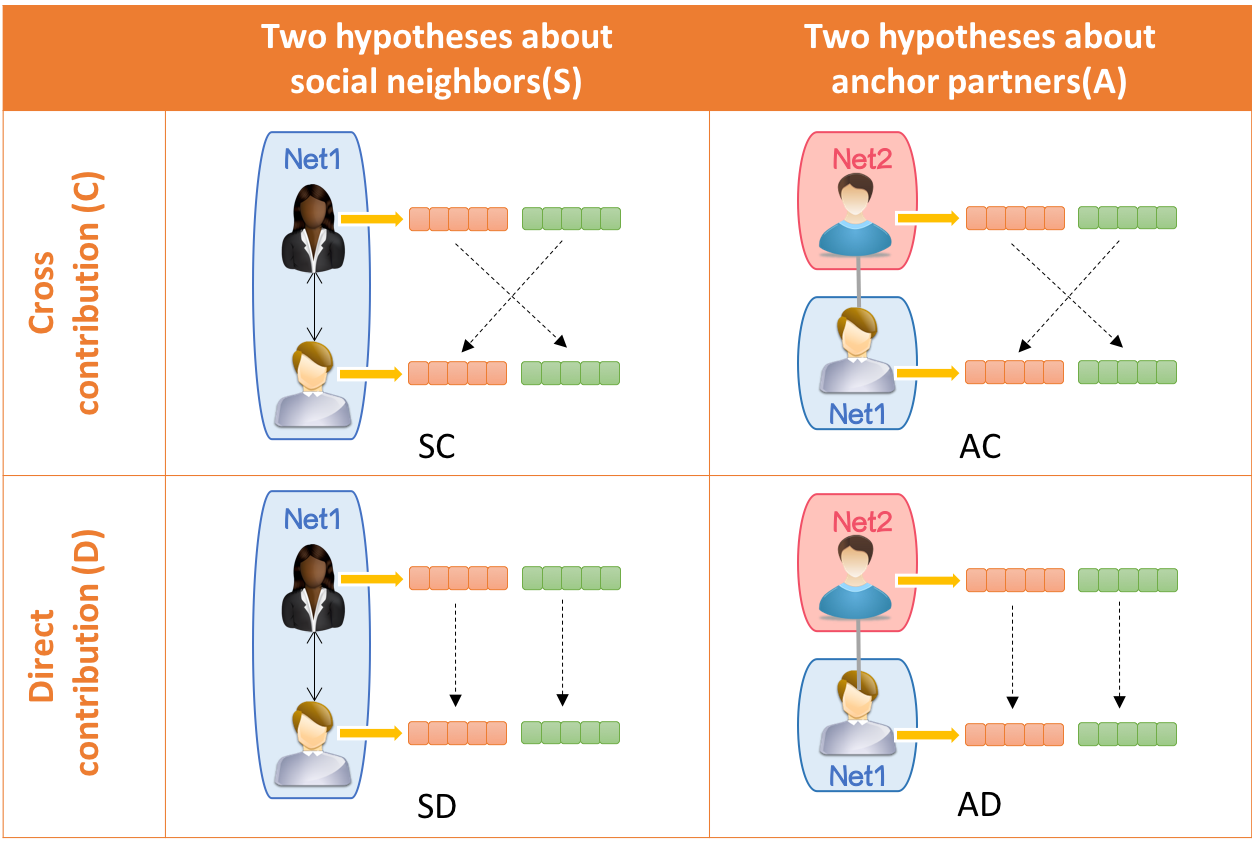}
     \caption{Four hypotheses about how the neighbors contribute to the target node. There are two kinds of contribution modes for social neighbors and anchor partners respectively.}
     \label{fig:correspondence}
     \vspace{-5pt}
 \end{figure}

As mentioned in the method part, in our framework, each node is represented with the initiator and recipient features and it is crucial to determine how to aggregate information from the neighbors connected by social links and anchor links. If the neighbor's initiator and recipient features contribute to the target node' s recipient and initiator respectively, we name it as cross-contribution. Conversely, if the initiator and recipient features of two neighbor nodes are related correspondingly, we name it direct-contribution. By combining two mechanisms with two kinds of links, there are four different hypotheses as illustrated in Figure \ref{fig:correspondence}. 

The hypotheses SC+AD is adopted in our framework. To validate it, we study the variants of our full model with the other three hypotheses and compared their performances on the collective link prediction task in Table \ref{tab:correspondence}. The experimental results indicate the model with SC+AD can achieve the best performance in both social link prediction and anchor link prediction. If direct-contribution is replaced with cross-contribution for anchor links, AUC of the anchor link prediction decreases a lot in two datasets. And the performances of the social link prediction are affected if we adopt direct-contribution for social links.

\begin{table}[!htbp]
\center
\caption{Hypothesis verification. SC+AD is adopted in our framework and achieves the best results.}
\label{tab:correspondence}
\begin{tabular}{c|ccc|ccc}
\toprule
  \multirow{2.5}*{Hypothesis}& \multicolumn{3}{|c|}{Twitter\&Foursquare}& \multicolumn{3}{|c}{Facebook\&Twitter}\\
  \cmidrule(lr){2-7}
  & Soc1 & Soc2 & Ach & Soc1 & Soc2 & Ach \\ 
\midrule
  SC+AC		&97.2\%	&96.0\%	&89.4\%	&98.1\%	&97.0\%	&96.0\%	\\
  SD+AD	 	&91.8\%	&91.3\%	&82.8\%	&97.0\%	&95.2\%	&95.3\%	\\
  SD+AC	 	&91.8\%	&91.3\%	&82.1\%	&96.9\%	&95.3\%	&94.3\%	\\
  \textbf{SC+AD}		&\textbf{97.2\%}	&\textbf{96.7\%}	&\textbf{92.4\%}	&\textbf{98.1\%}	&\textbf{97.1\%}	&\textbf{96.7\%}	\\
  \bottomrule
\end{tabular}
\vspace{-10pt}
\end{table}

\begin{figure}[!htbp]
  \centering
  \subfigure{
    \label{fig:test1}
    \includegraphics[width=0.47\linewidth]{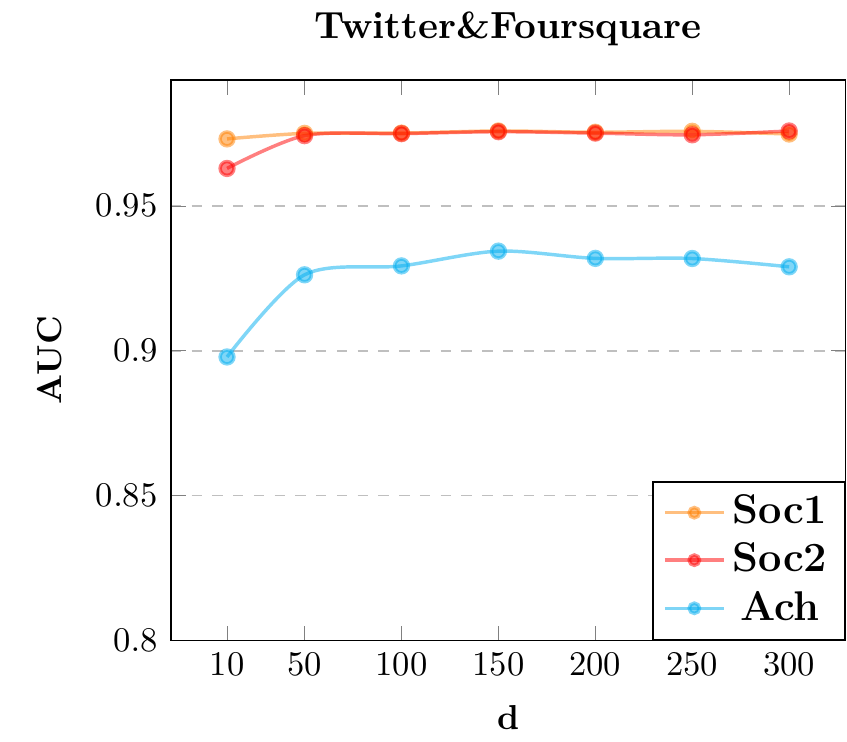}
 }
  \subfigure{
    \label{fig:test4}
    \includegraphics[width=0.47\linewidth]{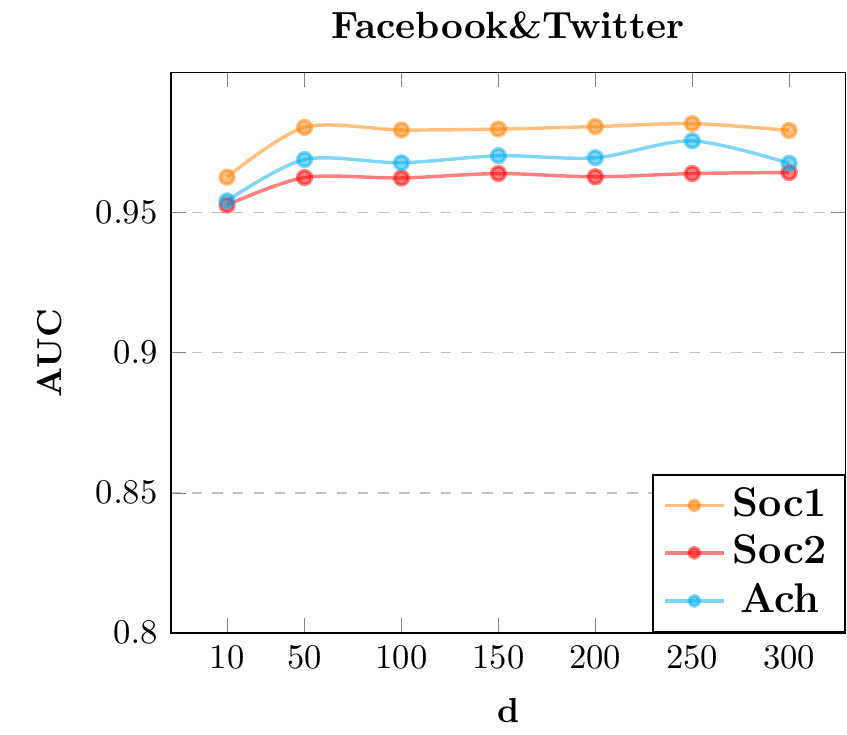}
 }
  \subfigure{
    \label{fig:test2}
    \includegraphics[width=0.47\linewidth]{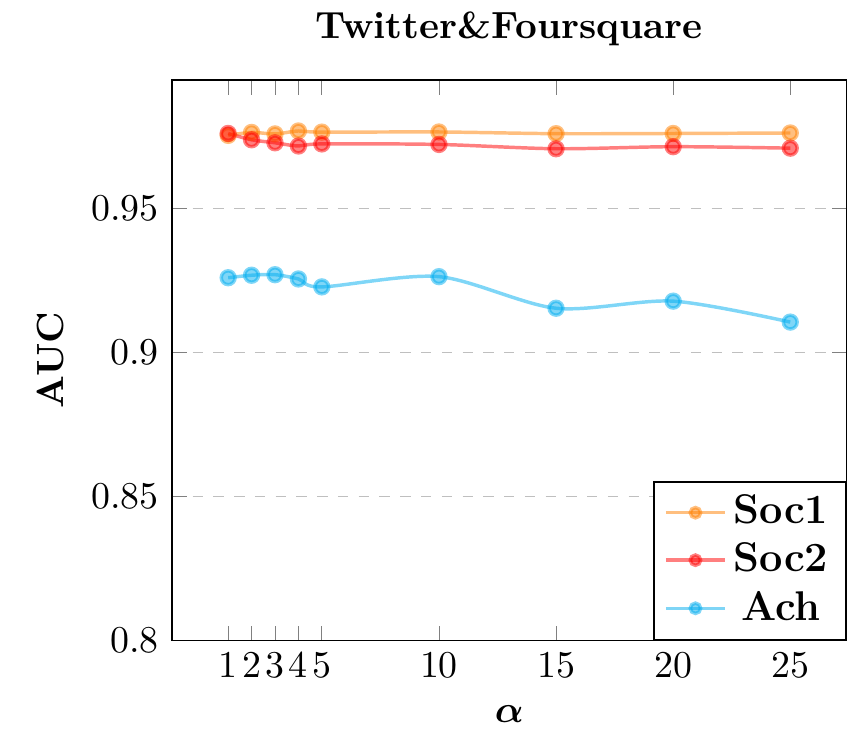}
 }
  \subfigure{
    \label{fig:test5}
    \includegraphics[width=0.47\linewidth]{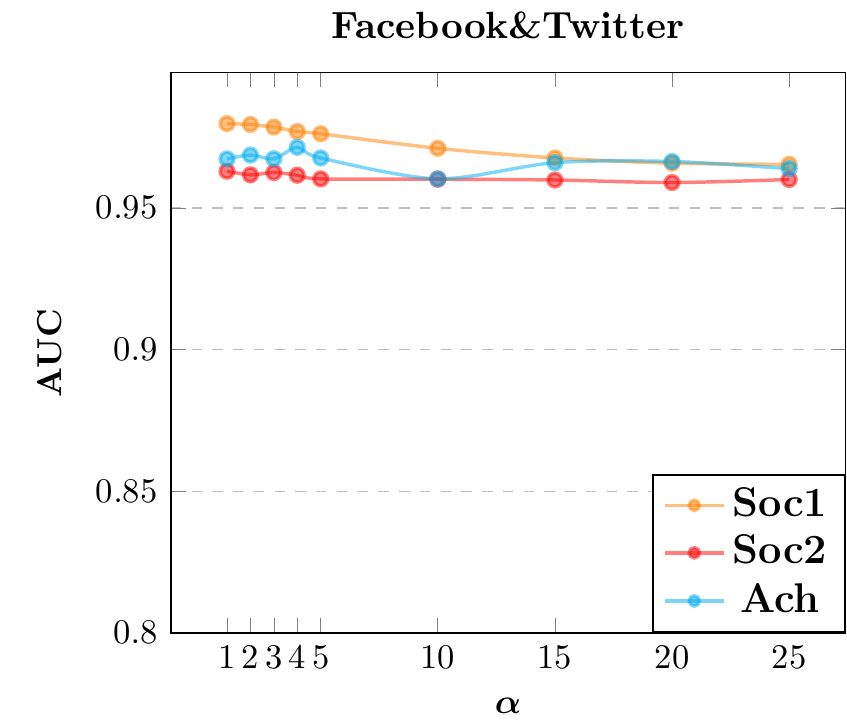}
 }
  \subfigure{
    \label{fig:test3}
    \includegraphics[width=0.47\linewidth]{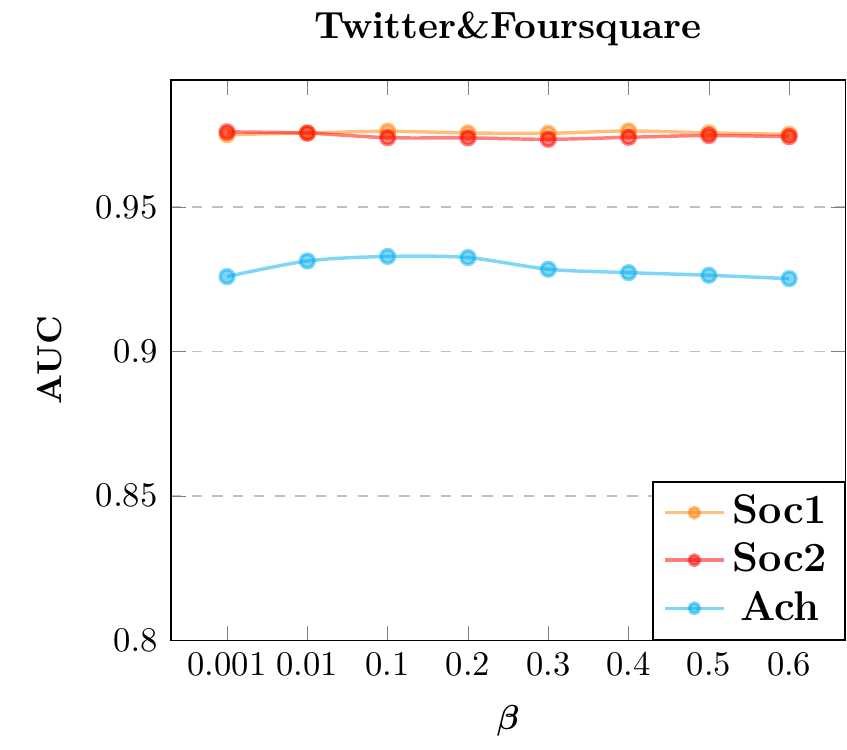}
 }
  \subfigure{
    \label{fig:test6}
    \includegraphics[width=0.47\linewidth]{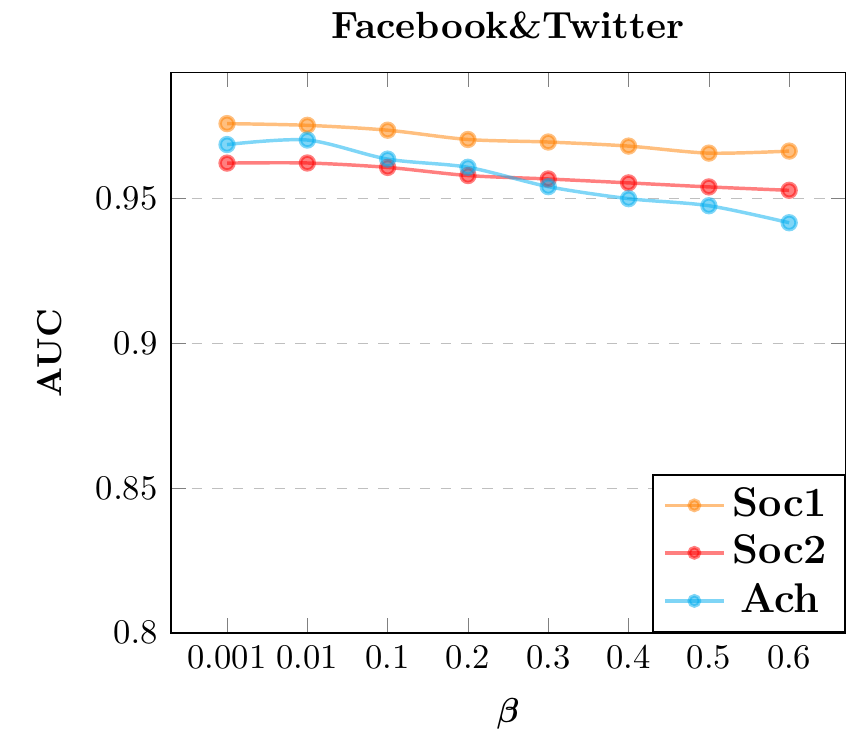}
 }
 \caption{Hyperparameter analysis. Our method is robust to choices of $d$, $\alpha$ and $\beta$ on two datasets}
 \label{fig:three-testing}
 \vspace{-10pt}
\end{figure}

\subsection{Parameter Analysis}

Now we examine the influence of three key parameters in our framework: the embedding size $d$ and the weight of the information transfer objective $\alpha$ and the weight of the regularization $\beta$. The three subfigures on the left in Figure \ref{fig:three-testing} show the sensitivity analysis on the first dataset while the rest is about the second dataset. The results in the first two subfigures indicate that setting the embedding size $ d $ to 100 can provide the best performance on both two datasets. Even when $ d $ is as low as 10, our model can achieve good results on three kinds of link prediction simultaneously. 

The parameter $\alpha$ denotes the strength of aligning the two networks. The two subfigures about $\alpha$ in the middle show how different values of $\alpha$ can affect the results on different datasets. The optimal $\alpha$ is near 1.0. When setting $\alpha$ in [1, 3], all the link prediction tasks perform well and stably. Anchor link prediction and social link prediction in sparser networks are affected as $\alpha$ increases. However, social link prediction in the dense networks is still stable. For the weight parameter $\beta$, the best setting is in [0.1, 0.2] according to the last two subfigures. It has a certain impact on anchor link prediction across the network while social link prediction within the network is not that sensitive to the parameter $\beta$ on both two datasets. 

\section{Related work}

\textbf{Multi-Network Analysis} Traditional network embedding methods focus on one single network \cite{perozzi2014deepwalk, grover2016node2vec, zhang2017learning} and suffer from the data insufficiency problem in the cold start scenarios. Therefore,  multi-network analysis has been a hot research topic and studied for data enrichment for several years on which dozens of works have been published \cite{zhang2018social, zhang2019broad}. Some work studied on information transfer across networks by anchor links to improving the quality of inter-network link prediction \cite{zhang2013predicting, zhang2014transferring, zhang2014meta, zhang2017bl, cao2018neural}. Besides, many existing works aim at anchor link formation prediction automatically \cite{zhang2015multiple, kong2013inferring}. However, most of these works study either intra-network or inter-network link prediction tasks separately. Zhang et al. first proposed the collective link prediction task \cite{zhan2018integrated}. The existing methods mostly ignore the contradiction of different characteristics of aligned networks or adopt fixed parameters to control the proportion of information diffused across networks, which usually need to be fine-tuned manually. Besides, these works also fail to consider the link connectivity of the links within and across networks.

\textbf{Neural attention mechanism} Neural attention mechanism has inspired many state-of-the-art models in several machine learning tasks including image caption generation \cite{xuk2015show}, machine translation \cite{vaswani2017attention, gehring2017convolutional} and semantic role labeling \cite{tan2018deep}. Its effectiveness is owed to making the model focus on more important detailed information and neglecting the useless information. In recent years, some works have also investigated the use of attention on graphs \cite{choi2017gram, DBLP:journals/corr/abs-1710-10903, zhang2018gaan}. Our work propose the hierarchical graph attention to model multiple aligned networks and overcome network characteristics contradiction to transfer more effective information across networks.

\section{Conclusion}
In this paper, we study the collective link prediction problem in multiple aligned social networks. We propose a novel application oriented network embedding framework, namely Hierarchical Graph Attention based Network Embedding ({\our}) to learn node embeddings. The hierarchical graph attention mechanism is introduced to resolve the network characteristic differences and link directivity differences. We conduct detailed empirical evaluations using several real-world datasets and the results demonstrate that our model outperforms other competitive approaches and handles the collective link prediction problems effectively.

%
\begin{acks}
This work is supported in part by the National Natural Science Foundation of China Projects No.U1636207 and NSF through grant IIS-1763365. This work is also partially supported by the Shanghai Science and Technology Development Fund No. 16JC1400801 and the National Natural Science Foundation of China Projects No.91546105.
\end{acks}

\bibliographystyle{ACM-Reference-Format}
\bibliography{sample-base}

\end{sloppypar}
\end{document}